# Object-Oriented Architecture: A Software Engineering-Inspired Shape Grammar for Durand's Plates

**Rohan Agarwal**
*Georgia Institute of Technology, USA*
roaga@gatech.edu

Addressing the challenge of modular architectural design, this study presents a novel approach through the implementation of a shape grammar system using functional and object-oriented programming principles from computer science. The focus lies on the modular generation of plates in the style of French Neoclassical architect Jean-Nicolas-Louis Durand, known for his modular rule-based method to architecture, demonstrating the system's capacity to articulate intricate architectural forms systematically. By leveraging computer programming principles, the proposed methodology allows for the creation of diverse designs while adhering to the inherent logic of Durand's original plates. The integration of Shape Machine allows a flexible framework for architects and designers, enabling the generation of complex structures in a modular fashion in existing CAD software. This research contributes to the exploration of computational tools in architectural design, offering a versatile solution for the synthesis of historically significant architectural elements.

## Introduction

As opposed to architecture that is purely free-form and artistic, envisioned by a creative mind and sketched out meticulously by a creative hand, some in the field have envisioned something more structured. What if there was a recipe to follow to guarantee effective and beautiful design? What if with these recipes, an architect could experiment with a variety of configurations, plans, and details on a whim? This is the promise of this alternate approach to architecture.

This promise is not a new one. Most recently, new design technologies such as CAD and AI assistance are enabling architects to design bigger, better, and faster. But going back a few decades, the field of shape grammars has offered a paradigm to generate designs modularly from a set of rules. [2] And going back a few centuries to early 19th century France, we once again see this promise in the theories of Jean-Nicolas-Louis Durand and his "procedure to be followed in the composition of any project." [1] There are many other examples of these ideas throughout the history of architecture and design.

This study explores a combination of these sources in order to identify effective approaches for beginning to realize this promise, and to evaluate its potential and current limitations. In particular, we create a visual, shape-based rule set in Shape Machine, a new shape grammar interpreter that works with shapes drawn in Rhino3D, a popular CAD software. The aim of this rule set is to replicate the core examples of Durand's system, that claims to generate any project (in his approved style), and expand upon them by automatically generating variations. To do so effectively, we will translate computer programming concepts such as object-oriented and functional programming into scripting with shapes. Through this case study, we explore the effectiveness of object-oriented design, of Shape Machine, and of Durand's design rules—all together. From there, we will have a clearer idea of that path forward in realizing this promise of true generative architectural design.

## Background

### Durand's Theory

Jean-Nicolas-Louis Durand was a French architecture teacher who compiled his lectures on his theory into his book, the *Précis of the Lectures of Architecture*. [4] He "regarded the Précis of the Lectures on Architecture (1802–5) and its companion volume, the Graphic Portion (1821), as both a basic course for future civil engineers and a treatise. Focusing the practice of architecture on utilitarian and economic values, he assailed the rationale behind classical architectural training: beauty, proportionality, and symbolism. His formal systematization of plans, elevations, and sections transformed architectural design into a selective modular typology in which symmetry and simple geometrical forms prevailed. His emphasis on pragmatic values, to the exclusion of metaphysical concerns, represented architecture as a closed system that subjected its own formal language to logical processes." [4] In other words, Durand abandoned all the artistic flair of old architecture and created systematic, modular, logical, pragmatic, and functional plans. They were still aesthetically pleasing, but they were produced through strict procedures and rules that defined his unique brand of French Neoclassical architecture.

What is this "procedure to be followed in the composition of any project"? From Durand's *Précis*, the steps can be summarized as starting with the whole and going to the individual parts, as follows [1]:

1. Understand the requirements of the whole building: what it is to be used for and what characteristics it should have.
2. Consider the structure of the whole building: one solid mass vs. interrupted by courtyards, number of stories of different blocks, etc.
3. Identify which rooms must exist and the relationship between their importance, position, and size.
4. For each kind of room, identify details: vaults, columns, etc.
5. Divide the space up into a grid based on the size of the rooms (the number of interaxes).
6. Finish the sketch with ornamentation as you see fit.

Modularity, logic, and systems—for any computer scientist, the potential parallels with modern technology are obvious. On a theoretical level, Durand's approach to architecture seems like it could be codified in computer code. Already, these steps can be read as pseudocode you might see for an algorithm.

Durand also includes visual examples of his plates in the Graphic Portion of his lectures. [1] Throughout his writing, he most refers to his plates 4 and 5, which can be considered as the core examples of his architectural theory. As such, we focus on these two plates and the variations presented with them in this study, and they are presented in Fig. 1 and Fig. 2.

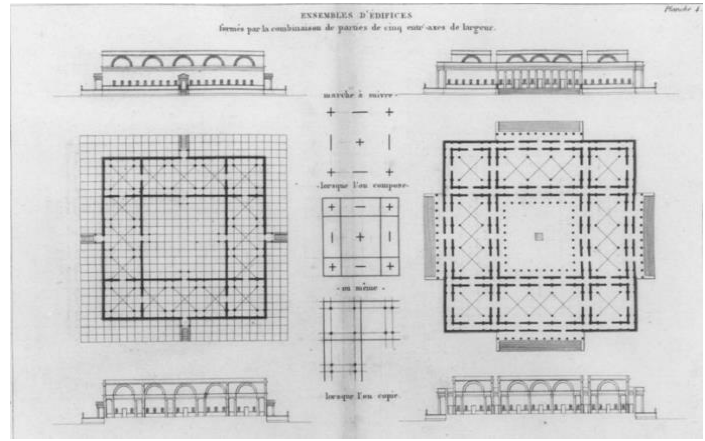

**Fig. 1** Durand's plate 4 from the Graphic Portion of the Lectures on Architecture [1]

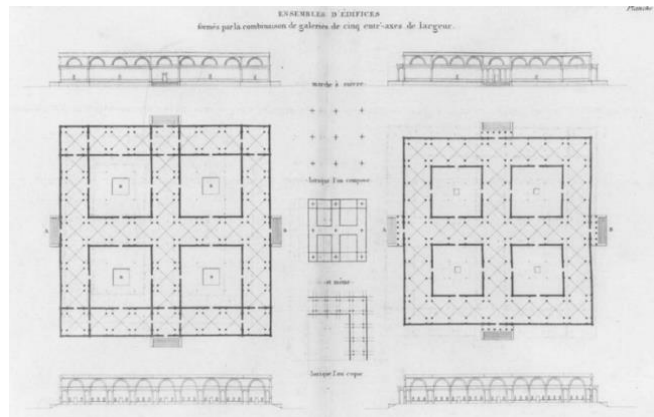

**Fig. 2** Durand's plate 5 from the Graphic Portion of the Lectures on Architecture [1]

Outlined in the center of both plates, Durand shows how he first marks the boundary of the design, then lays out the rooms, then begins on details of the columns using axes. Plate 4 is designed around a central courtyard and uses simple vaults and columns, and its variation on the right introduces corridors and other variations on ornamentation, such as in its stairs and courtyards. Plate 5 is designed around four courtyards and varies with the presence of walls in some areas and also with variations on ornamentation. Both plates use the same set of room shapes/sizes, just with different arrangements and detailing. [1] Overall, Durand outlines and demonstrates an approach to generate a variety of designs from his theoretical procedure, at least when drawing by hand, but potentially also when automated through shape grammars, computer programming, or a blend of the two.

**Shape Grammars and Shape Machine**

Shape grammars formalize the relationship between shapes. From a set of shape rules, that turn one shape into another shape, we get a system to generate a whole language of possible designs starting from a single initial shape, depending on how the rules are applied. [2] A simple example of this is shown in Fig. 3.

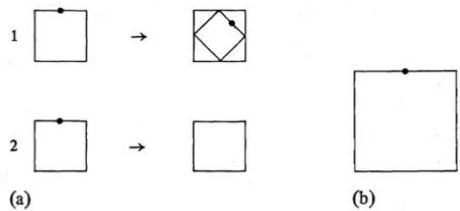

Figure 1. A simple shape grammar that inscribes squares in squares. (a) Shape rules, (b) initial shape.

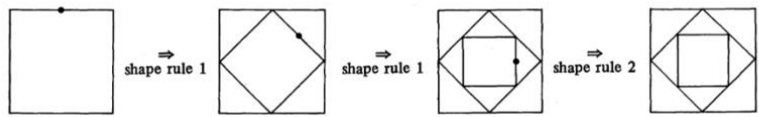

Figure 2. Generation of a shape using the shape grammar of figure 1.

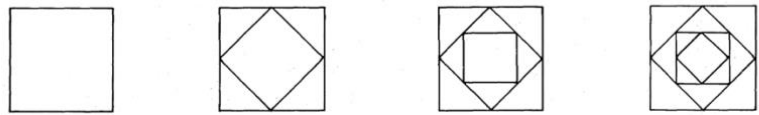

Figure 3. Some shapes in the language defined by the shape grammar of figure 1.

**Fig 3.** An example of a shape grammar, the process of applying shape rules to generate a design, and a language of possible designs from this grammar. [2]

Shape Machine is a software plugin for CAD software Rhino3D that lets designers draw and apply shape rules such as these. [3] Instead of using the raw curves in the CAD software, it can recognize shapes visually, such as the triangle formed by an inscribed square Fig. 3, making it more intuitive and powerful. In addition, Shape Machine features DrawScript+, which lets designers put these rules together in an executable program. DrawScript+ is Turing-complete, meaning it can theoretically solve any computation problem that another language can (though practical limitations may get in the way). It also allows for the creation of reusable "blocks" of code, combining them like functions or classes in programming, adding loops and if-then logic, and more. [5] This is a much more powerful application of shape grammars.

**Core Programming Concepts**

Considering that we now have a theoretical procedure for architectural design, a formalism for turning that into a generative program, and a software tool that can execute on that formalism in the form of CAD and proper Turing-complete computer programs, we are beginning to approach architecture as computer programming. Therefore, it is crucial to understand the fundamental paradigms used to write good programs.

The first paradigm is imperative programming. Here, a program is a series of commands (do this, do that). [7] Low-level languages such as Assembly are imperative. This paradigm is the most basic, but is not as easily expressive as others. A pure shape grammar (a sequence of rules) can be considered imperative.

Another paradigm is functional programming. Here a program is composed entirely of self-isolated functions, with no side effects. Functions can even be treated as variables or parameters in other functions. Such programs have no shared state. This paradigm is great for cleanliness, proving correctness, code reuse, and thread safety. [7]

The most popular paradigm in real-world programming is object-oriented programming (OOP), revolving around the concept of objects or classes that contain data and their own procedures that operate on that data. [6] Major principles of OOP include encapsulation (self-contained modules that keep data and methods private), inheritance (objects in hierarchies, allowing both code reuse and extending one object into multiple variations with unique behaviors; this opposes composition from functional programming), polymorphism (allowing any subclass to be interchangeable in a part of the program), and abstraction (representing the conceptual meaning of the object to its user; encapsulation hides all the details and abstraction is what faces externally). [6]

**Writing the Shape Grammar**
This study was performed with the Shape Machine plugin in Rhino 7. A grammar was created using Shape Machine's DrawScript+ mode, allowing for full scripting/programming capabilities.

**Generating the Whole**

As Durand outlined in his universal procedure, the first step is to understand the structure of the whole building. As such, the first step of the grammar will be to generate the general layout of the structure.

Looking closely at plates 4 and 5 in Fig. 1 and Fig. 2, both are constructed of the same "building blocks": an 11x11 courtyard surrounded by 5x5 and 5x11 rooms (units are relative). Considering that, we can follow a simple procedure starting with just one square.

It begins by separately marking the square in the horizontal direction and the vertical direction. It then loops vertically a set number of times, building out courtyards spaced properly. It then loops horizontally, constructing courtyards vertically spaced from all existing courtyards. After some cleanup and filling in missing walls, we jump to a separate routine that "breaks walls." A full step-by-step example is shown in Fig. 4. To break walls, we leverage Durand's principle of understanding the positional and size relationship of rooms in the layout. Each rule identifies a section based on this relationship of neighboring rooms and removes the wall segment. A user can add in any rules wanted to remove internal walls from the layout. Some examples are shown in Fig. 5, but any others can easily be added or removed.

A user of this grammar simply has to change the number of loops to perform in two rules: the horizontal construction and the vertical construction. This method allows us to generate versions of plates 4 and 5 of any dimensions (1x1 courtyard, 2x2 courtyards, 2x4, 3x11, or anything else). It does not generate L-shaped layouts, as that is out of scope for this study and later parts of the grammar, but the high-level approach in this section would still apply with some modifications. Only a few extra rules would be required in the later sections of the grammar, making it an easy extension.

This section is not necessarily required. A user can rapidly copy paste and stick together the three building blocks into a custom layout in Rhino. However, it demonstrates that a simple function can be written to automate everything. To automate even further, we could take advantage of the concept of polymorphism. We can create multiple wall-breaker functions, and choose between them based on a marker in the original design.

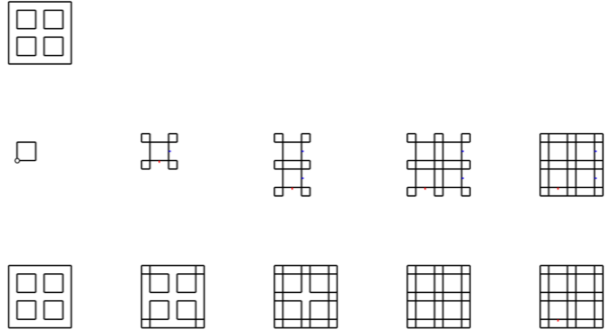

**Fig. 4:** An example of a 2x2 initial shape for plate 5 being constructed vertically, then horizontally, being filled in, and then walls being broken down.

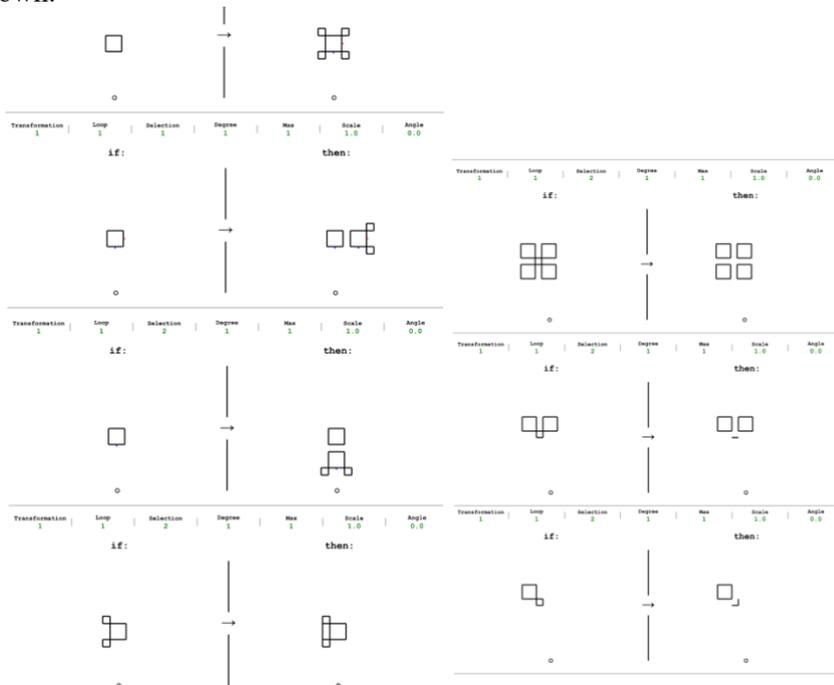

**Fig 5:** The core loop of expanding in two directions using markers (left) and example wall-breaker rules using positional and size relationships of rooms (right). Cleanup rules are omitted.

**Filling in the Parts**

Once a satisfactory layout whole is constructed, Durand's procedure asks us to fill in the details: doorways, columns, vaults, and ornamentation. This is where plates 4 and 5 and their variations branch out, requiring flexible, modular, reusable, and extensible code.

Here we must consider the paradigms of programming more thoroughly. While assembling rooms together makes object-oriented programming seem intuitive, Shape Machine does not support anything like it. We cannot have internal data or methods in Shape Machine. All shape rules operate on a shared canvas that acts as both input and output. And once an object (say a room) is created, it cannot be modified, moved, or referenced in any way in isolation. Instead, Shape Machine operates more naturally under the functional programming paradigm—functions calling and chaining other functions (rule blocks/sets). However, unlike functional programming, all functions have a global state, or shared canvas. Considering this in-between nature of Shape Machine, we will still need to leverage OOP concepts of encapsulation, abstraction, inheritance, and polymorphism to design our functions as flexibly as possible. This also helps in composing them intuitively, like a factory assembly line with a kit of parts.

We will first create a grammar for Plate 5, and then make the needed extensions and modifications for Plate 4. Finally, a general, overall grammar will be clear. Note that not every ornamentation detail in Durand's original plate was replicated in these examples due to time constraints, but it should be trivial to swap one design for another.

*Plate 5*

Similarly to the grammar generating the whole layout, this grammar continues considering the position, shape, and size relationships between neighboring rooms. With these relationships, the grammar can identify specific room types and give them a unique marker, so they can later be recognized and treated individually. It leverages functional composition to string together different relationships for different room types (5x5, 5x11, 11x11 and their variants, depending on neighbor relationship) under a single "room marker" function.

The main function begins with this room marker function and then deletes the layout sketch, leaving just the markers. It then uses a "room builder" function, which, similar the "room marker," composes detailing functions for each room variant. Details placed in the room builder include doorways, columns, vaults, stairs, and walls, some of which are marked uniquely for later processing. An example of the room marker and room builder can be found in Fig. 8.

Finally, the main function uses a "detailer" function to process the skeleton into the final plate. It details basic marker shapes, such as stairs and courtyards. It also shifts column markers on walls into piers. It also cleans up extraneous vault lines inside walls. The "detailer" function also is composed with another "wall builder" function that turns single lines that mark a wall location into a thick wall with depth.

A step-by-step example of this grammar's outputs can be found in Fig. 6. An overview of the whole grammar is presented in Fig. 7. Notably, combined with either the first grammar generating the layout or by manual assembly of the building block rooms, a great number of variations of plate 5 can be generated by this grammar. A few examples are demonstrated in Fig. 9. If a user wanted to support a new kind of room, they would simply have to add the corresponding marker and builder rules (or, to speak in terms of OOP, extend the existing class and swap it in with polymorphism). The grammar shown here provides an abstract interface that allows endless extension, substitution, and modularity—build the layout, mark the rooms, build the rooms, and detail the result.

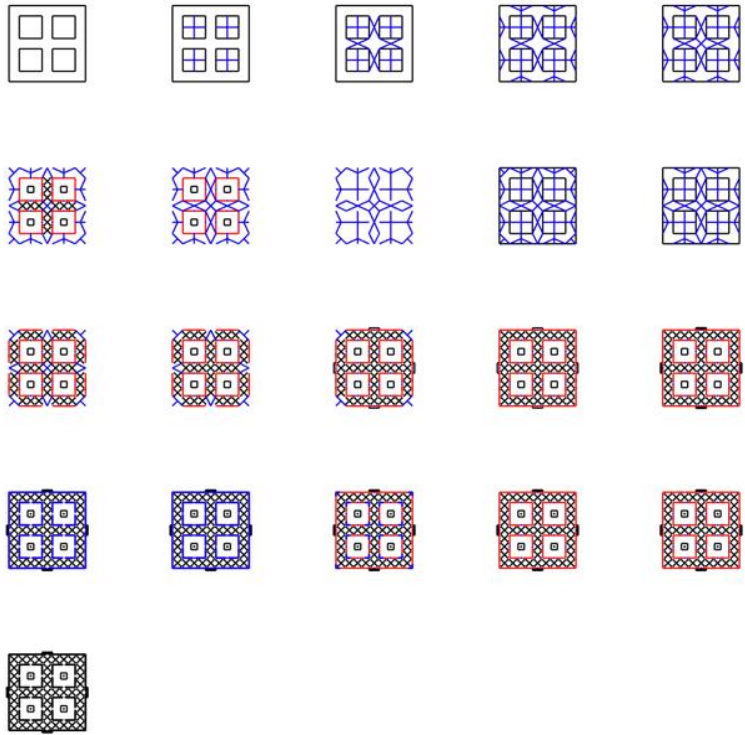

**Fig 6:** An example generation of plate 5, starting from a layout to the finished plate. It marks each room in blue by its relationship to its neighbors, removes the layout (black), and begins detailing (red and black). Finally, the red lines are turned into full walls.

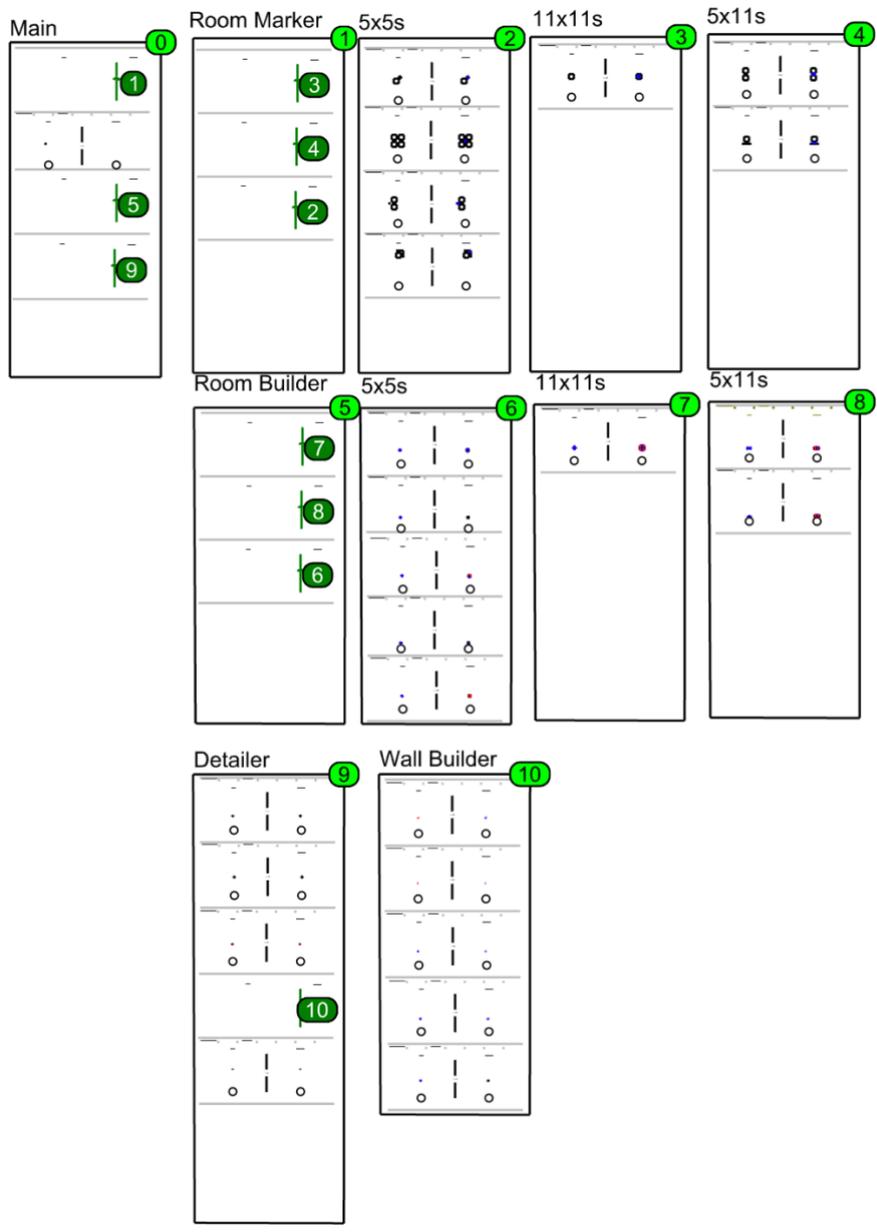

**Fig 7:** An overview of the whole grammar, with a room marker, room builder, and detailer with wall builder.

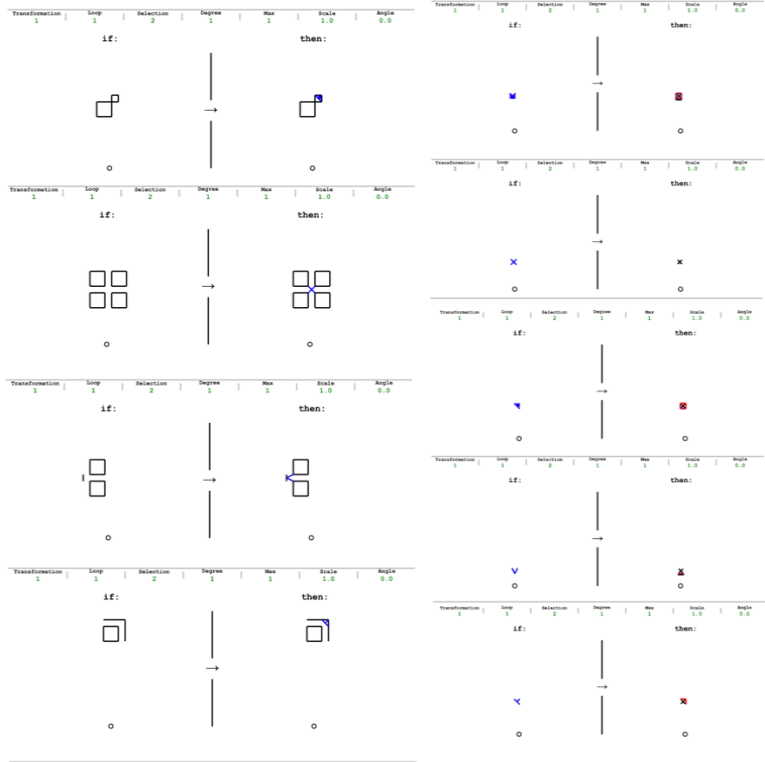

**Fig. 8:** An example of the room marker function for different 5x5 variants (left) and the room builder function that identifies those markers (right).

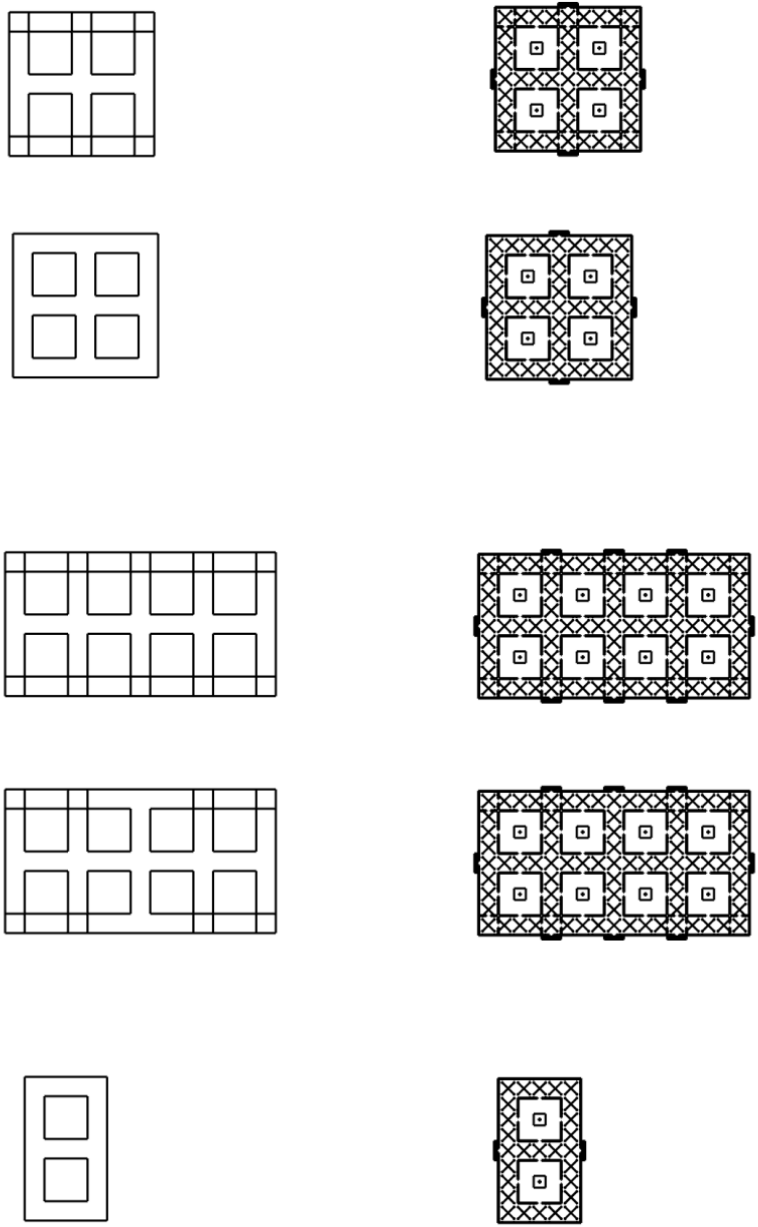

**Fig. 9:** Different possible variations of Durand's plate 5 achieved by this grammar.

*Plate 4*

In adapting this grammar to plate 4, the overall interface and hierarchy of the grammar remains the same. Only some rules in marking, building, and detailing were changed to match Durand's designs. These changes can be thought of in OOP terms as creating a new subclass of the 5x5/5x11/11/x11 or any other object, that by polymorphism, can be used equivalently with the existing abstract interface (mark, build, detail). This demonstrates the generality of this grammar.

Durand, however, introduced a design for corridors in his second version of plate 4, requiring large changes to the structure of a building. In order to preserve the reusability of as much code as possible, this grammar adds corridors after all rooms are marked and built, and before detailing is complete. This is done simply by composing another function, "offset corridors," inside the main function. In the current grammar, a user has to drag in or drag out the single jump rule in the main method to indicate if they want corridors or not, but this could easily be done with a visual indicator in the initial shape instead. A step-by-step example of this generation is presented in Fig. 10. Some more possible variations from this grammar are shown in Fig. 11. Note the version of plate 4 with a dome in Fig. 11, which uses a new dome-building function that is composed in the detailing function (in-place of the usual courtyard detail). This is just one example of how one can easily compose and plug-in additional modules/functions into the grammar to generate a variety of designs.

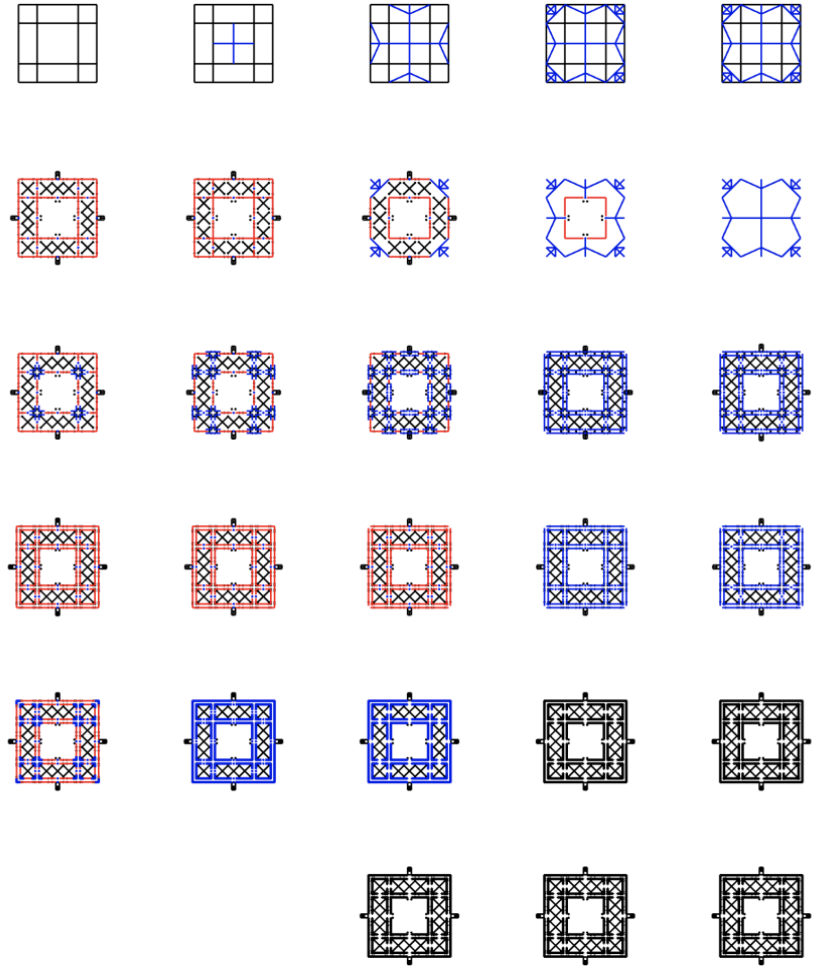

**Fig. 10:** A step-by-step example of this grammar generating a variant of plate 4 with corridors.

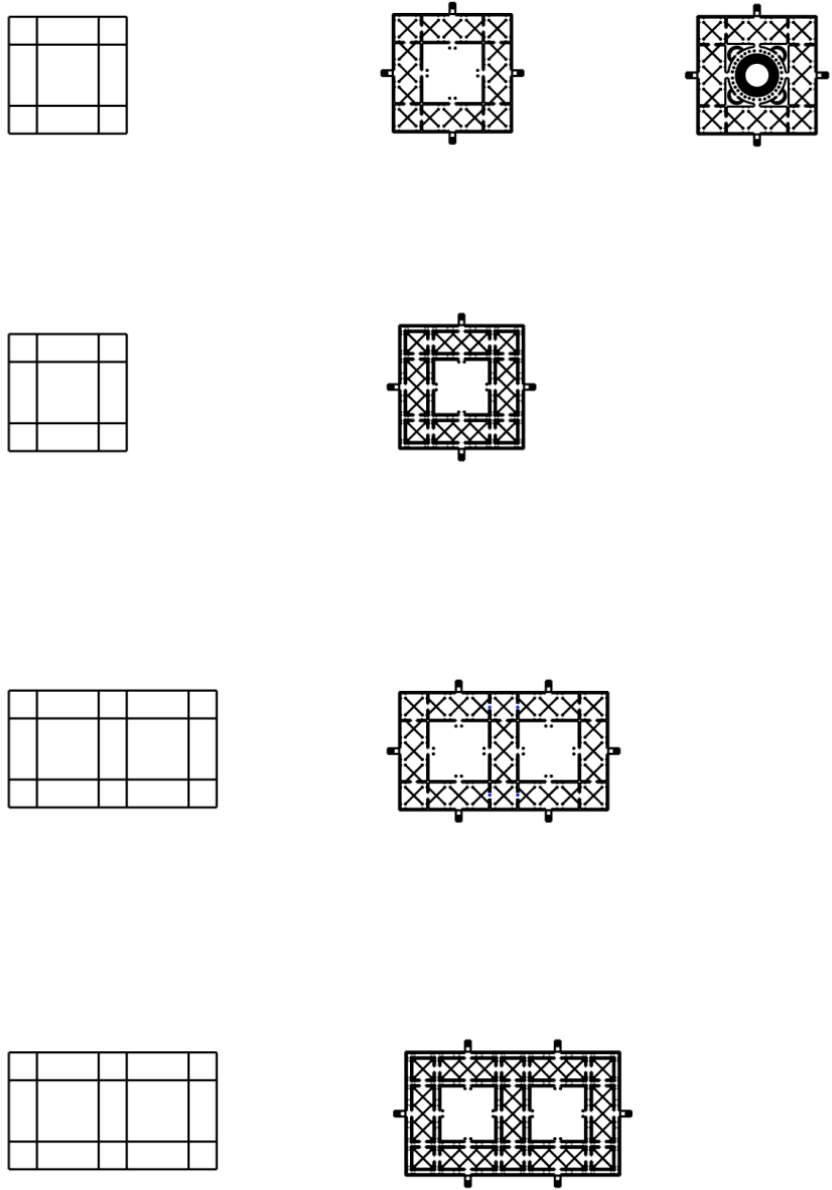

**Fig. 11:** Some different variations of plate 4 that can be generated with this grammar in Shape Machine.

## Discussion

The ability to create variations in structure and layout has been demonstrated. The ability to create variations in details and design, such as with domes and courtyards, has also been demonstrated. Much of this was achieved with no change to the overall structure and abstract interface of the grammar, and minimal changes to other code. The grammar leveraged both object-oriented programming concepts and function composition techniques.

However, can we combine these into one grammar? If we manually drag and drop out rules, we can accomplish this trivially (and tediously). The better approach is to maintain various separate functions for different plates as needed, and compose or swap them out by simply changing the number in the JUMP instruction. This approach applies to marking, building, and detailing. It leverages a constant interface, the idea of polymorphism, and a vague idea of inheritance and subclasses. From the current grammar, this hierarchy might need to be introduced in a few places to clean them up, such as the corridors function.

Sometimes, though, that approach would require changing a few JUMP instructions at once—can we accomplish this without any code modification at all? We can make JUMPs conditional on a marker on the canvas, treating the initial shape as also a user configuration of which variant to select. And with Shape Machine's abilities to make random selections, the grammar could be adapted to give a new variation every time it runs.

How far can we generalize this grammar? With these approaches, it could make any number of variations in theory. The class/function hierarchy and rule sets—meaning which details, room sizes, and room relationships are supported--may just have to grow with them, as well as program runtime for complex designs. Still, this interface is a great simple start, as long as you have a layout of building block rooms ready.

## Conclusion

This grammar is a generalizable, modular system to generate a variety of designs in the style of Durand. It translates Durand's proposed procedure, yet showcases the value of modern functional and object-oriented programming concepts in the world of shape grammars and architectural design. Future work could be done to make this grammar endlessly more generative with more rules and variations supported, and with other user interface considerations in specifying design decisions. As Shape Machine and DrawScript+ develop, more programming patterns and paradigms could also be introduced to simplify the code or make it more powerful. An investigation into best practices for shape grammar performance/efficiency would also be valuable.